\documentclass[aps,twocolumn,groupedaddress,showpacs]{revtex4}
\usepackage{graphicx}
\begin{document}

\title{Dynamical structure factors of $S=1$ bond-alternating Heisenberg chains }

\author{Takahumi Suzuki and Sei-ichiro Suga}
\affiliation{Department of Applied Physics, Osaka University, Suita, Osaka 565-0871, Japan}
\date{\today}
\begin{abstract}
We calculate the dynamical structural factor of the $S=1$ bond-alternating Heisenberg chain. 
In the Haldane phase, the lowest excited states form the lower edge of the multimagnon continuum in $0 \leq q \leq q_c$ and the one-magnon mode in $q_c \leq q \leq \pi$. As the system approaches the gapless point, $q_c$ shifts towards $q=\pi$ and the largest integrated intensity of the one-magnon mode is decreased.  
In the singlet-dimer phase, the one-magnon mode appears in $0 \leq q \leq q_c$. As the bond-alternation becomes strong, $q_c$ shifts towards $q=\pi$. 
In the antiferromagnetic-ferromagnetic bond-alternation region with a strong ferromagnetic coupling, the lowest excited states form the lower edge of the multimagnon continuum in $0 \leq q \leq 0.2\pi$ and $0.8\pi \leq q \leq \pi$, and the one-magnon mode appears in $0.2\pi<q<0.8\pi$. 
The largest integrated intensity of the one-magnon mode is $93\%$, which is slightly smaller than that in the $S=1$ Haldane-gap system. 
We further discuss the dynamical structural factor in connection with the inelastic neutron-scattering experiments. 
\end{abstract}
\maketitle
%

\section{INTRODUCTION}
 The isotropic Heisenberg chain with integer spins has a finite excitation gap above a disordered ground state, while that with half-odd-integer spins has a gapless excitation from a critical ground state \cite{hal1,hal2}. 
This Haldane conjecture based on the O(3) nonlinear $\sigma$ model with a topological term has been confirmed by extensive numerical and experimental studies. 
The elementary excitations of the $S=1$ isotropic Heisenberg chain were investigated by inelastic neutron-scattering experiments. Using $S=1$ quasi-one-dimensional Heisenberg antiferromagnets ${\rm Ni(C_2H_8N_2)_2NO_2ClO_4}$ \cite{ma,zal} and ${\rm CsNiCl_3}$ \cite{ken1,ken2}, the dynamical structure factors $S(q,\omega)$ for the one-magnon mode and the multimagnon continuum were observed. The integrated $S(q,\omega)$ with respect to $\omega$ at $q=\pi$ from the multimagnon mode was compared with the results obtained by several theoretical methods \cite{gom,wh,taka,kw,ess,ha}. The origin of the deviation was discussed in connection with the picture for the elementary excitation \cite{ken2}. 
 
 After some years later from the Haldane conjecture, Affleck \cite{aff1,aff2} and Affleck-Haldane \cite{ah} further applied the same analysis to the bond-alternating spin-$S$ Heisenberg chain described by the following Hamiltonian: 
%
\begin{eqnarray}
H= J\sum_{i} \Biggl(\mathbf{S}_{2i-1}\cdot\mathbf{S}_{2i} + 
   \alpha\mathbf{S}_{2i}\cdot\mathbf{S}_{2i+1}\Biggl), 
\end{eqnarray}
%
and argued that there exist $2S$ gapless points in $0 < \alpha < \infty$. Their argument was successfully confirmed by several theoretical methods \cite{sg,kt,ys1,ys2,totsuka1,yt,KN,kohno,NT}.  For the $S=1$ system, the critical $\alpha$ separating the Haldane phase $(\alpha_c<\alpha<1)$ and the singlet-dimer phase $(0<\alpha<\alpha_c)$ was estimated using a series expansion method \cite{sg}, a density-matrix renormalization-group method \cite{kt}, quantum Monte Carlo methods \cite{ys1,ys2,kohno,NT}, the Binder parameter combined with finite-size scaling \cite{totsuka1}, and a level-crossing method \cite{KN}. The results are in agreement as $\alpha_{\rm c} \sim 0.6$. 
When $-\infty < \alpha < 0$ in the $S=1$ system, on the other hand, the trivial dimer phase around $\alpha \sim 0$ is shown to be smoothly connected to the $S=2$ Haldane phase which appears in the limit $\alpha \rightarrow -\infty$ \cite{totsuka1}.

Experimentally, using some nickel chain compounds which are regarded as the $S=1$ bond-alternating Heisenberg system, magnetic susceptibility, magnetization, and ESR measurements were performed \cite{nakano,naru1,hagi2,kohno,naru2}. From the comparison between the experimental and numerical results, the strength of the bond-alternation $\alpha$ was evaluated and it was concluded that some compounds are in the singlet-dimer phase \cite{nakano,naru1,naru2} and the other compound is close to the gapless point \cite{hagi2,kohno}.

The elementary excitations of the $S=1$ bond-alternating Heisenberg chain were investigated by a quantum Monte Carlo method \cite{ys2} and an exact diagonalization method \cite{totsuka1}. From the dispersion relation of the low-lying excitation, it was shown that the gap opens at the center of the Brillouin zone for $\alpha>0$, while the gap opens at the boundary of the Brillouin zone for $\alpha<0$. 
In the Haldane phase around $\alpha=1$, the low-lying excitations are expected to be scattering states of the domain walls in the hidden antiferromagnetic ordering \cite{ys2}. Around $\alpha=0$, in contrast, the low-lying excitations are well described by the $S=1$ magnon \cite{totsuka1}. 
In spite of these studies, dynamical properties have not yet been fully investigated. A detailed investigation is desirable.

 In this paper, we investigate dynamical properties of the $S=1$ bond-alternating Heisenberg chain. Using a continued fraction method based on the Lanczos algorithm \cite{GB}, we systematically calculate the dynamical structure factor (DSF) in the Haldane phase and dimer phase including the regions $\alpha>0$ and $\alpha<0$.  
The DSF provides us with information about the intensity of the magnetic excitation as a function of energy-momentum transfer. Since the DSF can be observed by inelastic neutron-scattering experiments, a comparison between theoretical and experimental results is possible. 
The setup of the paper is as follows. 
In Sec. II, we briefly summarize the method for the numerical calculation. 
In Sec. III, we show the results for the distribution of the intensity as a function of energy-momentum transfer. Using finite-size effects \cite{taka}, we discuss whether the lowest excited states in given wave numbers form an isolated mode or a lower edge of the excitation continuum.  
Sec. IV is devoted to the summary of the paper.

\section{MODEL AND METHOD}\label{sec2}
 We consider the $S=1$ bond-alternating Heisenberg chain described by the Hamiltonian (1). 
We set that the total number of spins is $N=20$ at the maximum and $J=1$. We apply the periodic boundary condition.
The DSF can be expressed in the form of the continued fraction \cite{GB} as
%
\begin{eqnarray}
S^{\mu}(q,\omega)
&=& -\frac{1}{\pi} \Im 
\langle\Psi_{0}|S^{\mu}_{q}\frac{1}{z-H}S^{\mu}_{q}|\Psi_{0}\rangle  
\nonumber \\
&=& S^{\mu}(q) C^{\mu}(q,\omega) \hspace{5mm} (\mu=x,y,z) ,
\end{eqnarray}
\noindent
where $| \Psi_{0} \rangle$ is the eigenstate with the lowest eigenvalue 
$E_{0}$, $S^{\mu}_q=(1/\sqrt{N})\sum_{j}e^{iqj}S^{\mu}_{j}$ and 
$z=\omega+i\eta+E_{0}$ with $\hbar=1$. 
The lattice constant between neighboring two sites is set to unity. 
Therefore, $q=0.5\pi$ corresponds to the boundary of the Brillouin zone.  
In the expression (2), $S^{\mu}(q)$ is the static structure factor and 
$C^{\mu}(q,\omega)$ is represented in the form of the continued fraction, which can be calculated numerically by Lanczos algorithm. 
The total contribution of $C^{\mu}(q,\omega)$ for fixed $q$ is normalized to unity, because the following sum rule has to be satisfied: 
$S^{\mu}(q)=\int^{\infty}_{0} {\rm d}\omega S^{\mu}(q,\omega)$. 
Since rotational symmetry around $x$, $y$ and $z$ axes remains, 
$S^{x}(q,\omega)=S^{y}(q,\omega)=S^{z}(q,\omega)$. We thus calculate only 
$S^{z}(q,\omega) \equiv S(q,\omega)(S^{z}(q) C^{z}(q,\omega) \equiv S(q) C(q,\omega))$. 
  Instead of taking $\eta\to+0$, we set $\eta = 1.0 \times 10^{-2}$. 
  For finite-size systems, therefore, $C^{z}(q,\omega)$ consists of a finite number of Lorentzians. Following Ref. 9, we call the position of a Lorentzian and the integrated value of each Lorentzian with respect to $\omega$ as the pole and residue, respectively.

To discuss whether the lowest excited states in given $q$ form an isolated mode or a lower edge of the excitation continuum in the thermodynamic limit, we investigate the finite-size effects of the poles and their residues of the continued fraction $C^{z}(q,\omega)$ \cite{taka,ys}.  As reported in Refs. 9 and 30-32, a pole which belongs to an excitation continuum tends to have an appreciable size dependence on at least either its position or its residue, while a pole which belongs to an isolated mode hardly depends on the system size. 

Using these methods, we proceed to the numerical calculation for the DSF and discuss characteristics of the low-lying excitation.

%
\section{RESULTS}\label{sec3}
We first investigate characteristics of the lowest excited state in a given $q$.  In Fig. 1, the finite-size effects for the residues of the lowest excited states are shown in the extended zone scheme. Note that the finite-size effects only for the residues are presented, since the positions of the poles scarcely depend on the system size. 
We discuss the results, making reference to the typical size dependence of the excitation continua for the $S=1/2$ \cite{ys} and $1$ \cite{taka} isotropic Heisenberg chains. 

In Fig. 2, the DSF's of the Haldane phase and the dimer phase are shown in the extended zone scheme. The intensity of $S^z(q,\omega)$ is proportional to the area of the circle. The full circles represent the isolated mode and the gray circles represent the excitation continuum. 
The largest intensity in a given $q$ appears at the lowest excited state. 
In $\alpha>0$ it increases as $q$ approaches $\pi$, while in $\alpha<0$ it increases as $q$ approaches $0.5\pi$. 
%

\subsection{Haldane phase}
As $N$ increases, the residues at $\alpha=0.8$ and $0.7$ decrease in $0<q<0.5\pi$ and in $0<q \leq 0.8\pi$, respectively. 
The size dependence of the residues in both regions of wave numbers is quite similar to that for the excitation continua of the $S=1/2$ \cite{ys} and $1$ \cite{taka} isotropic Heisenberg chains. 
The behavior of the residue around $q=0.9\pi$ at $\alpha=0.7$ is different from that in the other wave numbers. Such size dependence is probably caused by the fact that the next lowest excited state around $q=0.9\pi$ lies close to the lowest excited state with almost the same scattering intensity. Therefore, the size dependence around $q=0.9\pi$ suggests that the lowest excited states may become the lower edge of the multimagnon continuum. 
Accordingly, at $\alpha=0.8$, the lowest excited states in $0 \leq q<0.5\pi$ become the lower edge of the multimagnon continuum, while those in $0.5\pi \leq q \leq \pi$ form the one-magnon isolated mode. 
At $\alpha=0.7$, the lowest excited states in $0 \leq q \leq 0.9\pi$ become the lower edge of the multimagnon continuum, while those around $q=\pi$ form the one-magnon isolated mode. 

In Fig. 2, we show the DSF at $\alpha=0.8$ and $0.7$ in the extended zone scheme. The largest intensity caused by the one-magnon mode appears at $q=\pi$. 
We evaluate its integrated intensity by extrapolating at $N \rightarrow \infty$. The ratio that the one-magnon mode carries at $q=\pi$ is $87\%$ in $\alpha=0.8$ and $74\%$ in $\alpha=0.7$, respectively. These values are smaller than that in the $S=1$ Haldane-gap system, $97\% \sim 98\%$ \cite{taka,kw,ha,ess}. 
Therefore, as the system approaches the gapless point in the Haldane phase, the one-magnon mode is reduced towards $q=\pi$ with decrease in the excitation gap and the scattering intensity at $q=\pi$.  

At $\alpha=0.6$, the system is in the vicinity of the gapless point \cite{sg,kt,ys1,ys2,totsuka1,KN,kohno,NT}. The residues in $0<q\leq\pi$ decrease with increasing $N$, indicating that the lowest excited states in  $0 \leq q \leq \pi$ form the lower edge of the excitation continuum. 

%
\subsection{Dimer phase}
Judging from the finite-size effects of the residues, the one-magnon isolated mode appears in $0 \leq q< 0.5\pi$ at $\alpha=0.5$, while the lowest excited states form the one-magnon mode in $0 \leq q \leq \pi$ at $\alpha=0.3$ and $0.1$. 
As the system leaves from the gapless point in the singlet-dimer phase, the wave-number region of the one-magnon mode seems to extend towards $q=\pi$. 
At $\alpha=-0.3$, the lowest excited states form the one-magnon mode in $0 \leq q<0.8\pi$ and the lower edge of the multimagnon continuum in $0.8\pi \leq q \leq \pi$. 

At $\alpha=-10.0$, the system can be regarded as the $S=2$ Haldane-gap system.  From the finite-size effects of the residues, we find out that the one-magnon mode appears in $0.2\pi<q<0.8\pi$, and the lowest excited states form the multimagnon continuum in $0 \leq q \leq 0.2\pi$ and $0.8\pi \leq q \leq \pi$. 
The integrated intensity of the one-magnon mode at $q=0.5\pi$ is evaluated to be $93\%$ by extrapolating at $N \rightarrow \infty$. 
The integrated intensity that the one-magnon mode carries at the boundary of the Brillouin zone in the $S=2$ Haldane-gap systems is smaller than that in the $S=1$ Haldane-gap system. 
%
%
\begin{figure}[htb]
\includegraphics[trim=0cm 0cm 0cm 0cm,clip,width=8cm]{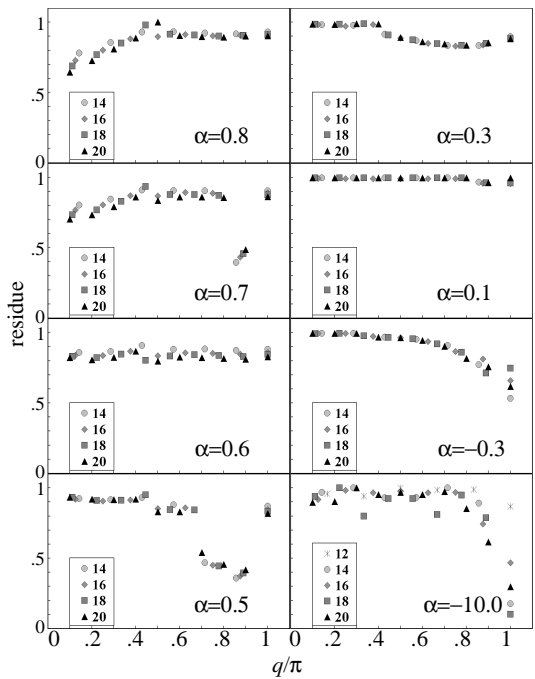}
\vspace{-1.5cm}
\caption{
The finite-size effects of the residues of the lowest excited states in given $q$.  
}
\label{Fig1}
\end{figure}
%

\subsection{Relation to inelastic neutron-scattering experiments}
We now investigate the observable DSF by inelastic neutron-scattering experiments.  The excitation energies of the lowest excited states are summarized in Table I.  Comparing the excitation energies at the symmetric wave numbers about $q=0.5\pi$, we discuss the DSF in the reduced zone scheme. 

At $\alpha=0.8$ and $0.7$ the excitation energies in $0.5\pi<q<\pi$ lie below the corresponding ones in $0<q<0.5\pi$. Therefore, at $\alpha=0.8$, the lowest excited states in $0 \leq q \leq 0.5\pi$ in the reduced zone scheme form the one-magnon isolated mode. At $\alpha=0.7$, the lowest excited states around $q=0$ form the one-magnon isolated mode, while those in $0.1\pi \leq q \leq 0.5\pi$ form the lower edge of the multimagnon continuum. 
The largest intensity caused by the one-magnon mode appears at $q=0$ with the integrated intensity $\sim 87\%$ in $\alpha=0.8$ and $\sim 74\%$ in $\alpha=0.7$. 

In the dimer phase, the excitation energies take the same values at the symmetric wave numbers about $q=0.5\pi$ within our numerical accuracy. 
Thus, at $\alpha=0.3$ and $0.1$, the lowest excited states in the reduced zone scheme form the one-magnon mode in $0 \leq q \leq 0.5\pi$. 
At $\alpha=0.5$ and $-0.3$, the one-magnon mode coincides with the lower edge of the multimagnon continuum. 
In this case, the one-magnon mode may become unstable and only the multimagnon continuum emerges. 
In fact, it was observed by inelastic neutron-scattering experiments that, when the one-magnon mode crosses the multimagnon continuum, the large intensity of the one-magnon mode becomes invisible in the multimagnon continuum \cite{ma,zal,ken1,ken2}. 
At $\alpha=0.5$, therefore, the lowest excited states form the lower edge of the multimagnon continuum in $0 \leq q \leq 0.5\pi$.
At $\alpha=-0.3$, the lowest excited states form the lower edge of the multimagnon continuum in $0 \leq q \leq 0.2\pi$ and the one-magnon mode in $0.2\pi<q \leq 0.5\pi$.

At $\alpha=0.3$ and $0.1$, the one-magnon modes corresponding to the symmetric wave numbers coincide with each other. At $\alpha=-0.3$, there exists such region of the wave numbers. In these cases, the intensity is described as the sum of two one-magnon modes. Therefore, the largest intensity does not always appear at $q=0$ or $0.5\pi$ in the reduced zone scheme.  
At $\alpha=0.3$ and $0.1$, the intensity of the one-magnon mode increases as $q$ approaches 0. At $\alpha=-0.3$, in contrast, the largest intensity of the one-magnon mode emerges at $q=0.4\pi$. 

At $\alpha=-10.0$, the multimagnon continuum appears in $0 \leq q \leq 0.2\pi$ and the one-magnon mode appears in $0.2\pi<q \leq 0.5\pi$ with the largest scattering intensity at $q=0.5\pi$. 
The wave-number region for the one-magnon mode is smaller than that in the $S=1$ Haldane gap system \cite{taka}.  
The largest intensity caused by the one-magnon mode appears at $q=0.5\pi$ with the integrated intensity $93\%$. 

The DSF's in the reduced zone scheme are presented in Fig. 3. 
Experimentally, thermodynamic properties of the $S=1$  bond-alternating Heisenberg chain have been investigated so far. 
To obtain detailed information on the elementary excitation, the study of dynamical properties is needed. Such studies should shed more light on understanding the nature of the $S=1$  bond-alternating Heisenberg chain. 

%
%
\begin{figure}[hbt]
\includegraphics[trim=0cm 0cm 0cm 0cm,clip,width=8cm]{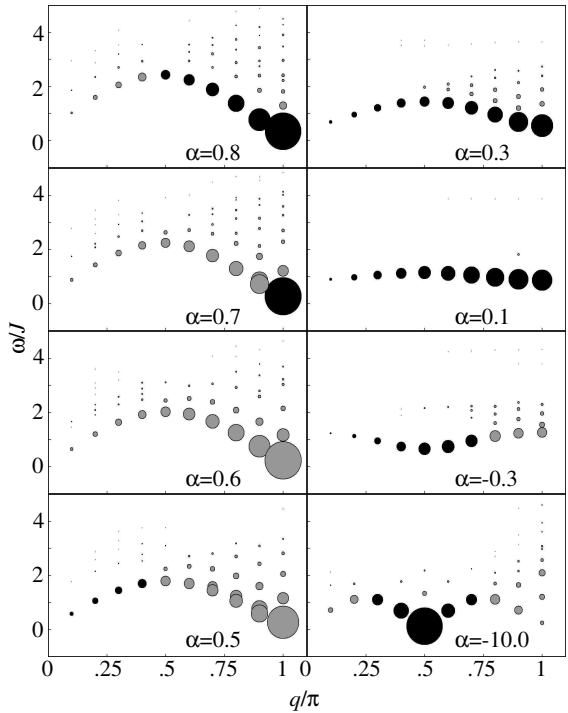}
\vspace{-1.5cm}
\caption{
$S(q.,\omega)$ for $N=20$ in the extended zone scheme. The intensity is proportional to the area of the circle. The full circles represent the isolated branch and the gray circles represent the excitation continuum.
}
\label{Fig2}
\end{figure}
%
%
\begin{figure}[hbt]
\includegraphics[trim=0cm 0cm 0cm 0cm,clip,width=8cm]{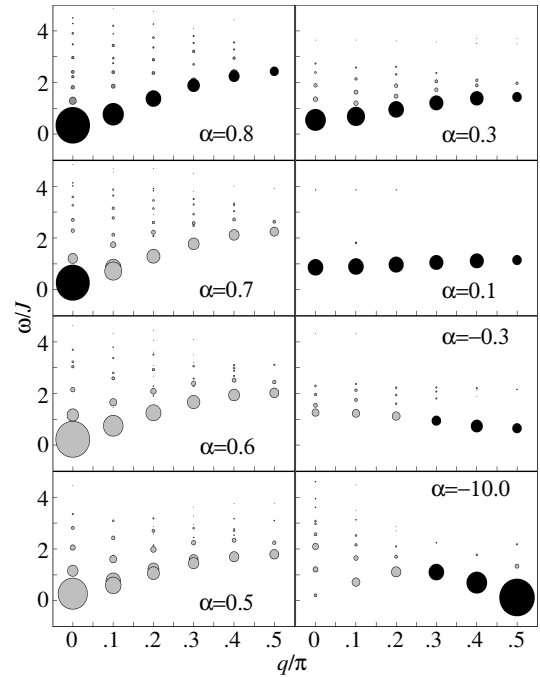}
\vspace{-2cm}
\caption{
$S(q.,\omega)$ for $N=20$ in the reduced zone scheme. The intensity is proportional to the area of the circle. The full circles represent the isolated branch and the gray circles represent the excitation continuum.
}
\label{Fig3}
\end{figure}
%
\section{SUMMARY}\label{Summary}
In this paper, we have calculated the DSF of the $S=1$ bond-alternating Heisenberg chain using a numerical diagonalization method. 
In the Haldane phase, the lowest excited states form the lower edge of the multimagnon continuum in $0 \leq q \leq q_c$ and the one-magnon mode in $q_c \leq q \leq \pi$. As the system approaches the gapless point, $q_c$ shifts towards $q=\pi$ and the largest integrated intensity of the one-magnon mode is reduced.  
In the singlet-dimer phase, the one-magnon mode appears in $0 \leq q \leq q_c$. As the bond-alternation becomes strong, $q_c$ shifts towards $q=\pi$. 
At $\alpha=-10.0$, the lowest excited states form the lower edge of the multimagnon continuum in $0 \leq q \leq 0.2\pi$ and $0.8\pi \leq q \leq \pi$, and the one-magnon mode appears in $0.2\pi<q<0.8\pi$. 

We have further discussed the DSF in the reduced zone scheme in connection with the inelastic neutron-scattering experiments.

\section{Acknowledgments}
We would like to thank K. Kakurai for fruitful discussions. 
Our computational programs are based on TITPACK version 2 by H. Nishimori. 
Numerical computations were carried out at the Yukawa Institute Computer Facility, Kyoto University, and the Supercomputer Center at the Institute for Solid State Physics, University of Tokyo. 
This work was supported by a Grant-in-Aid for Scientific Research from the Ministry of Education, Culture, Sports, Science, and Technology, Japan.


%
%
\begin{table}[htb]
\caption{
The excitation energies of the lowest excited states in the extended zone scheme for $N=20$. The boundary of the Brillouin zone is $q=0.5\pi$. 
}
\begin{tabular}{ccccccccccc} \hline \hline 
$\alpha$ & 
$q=\pi/10$ & 
$2\pi/10$ & 
$3\pi/10$ & 
$4\pi/10$ & 
$5\pi/10$ & 
$6\pi/10$ & 
$7\pi/10$ & 
$8\pi/10$ & 
$9\pi/10$ & 
$10\pi/10$ \\ \hline
$0.8$ & 
1.02 & 
1.59 & 
2.05 & 
2.35 & 
2.43 & 
2.24 & 
1.88 & 
1.37 & 
0.77 &
0.34 \\ 
$0.7$ & 
0.87 & 
1.42 & 
1.86 & 
2.15 & 
2.24 & 
2.11 & 
1.77 & 
1.29 & 
0.71 &
0.27 \\ 
$0.6$ & 
0.64 & 
1.19 & 
1.63 & 
1.92 & 
2.02 & 
1.94 & 
1.67 & 
1.25 & 
0.75 &
0.22 \\ 
$0.5$ & 
0.58 & 
1.06 & 
1.45 & 
1.70 & 
1.79 & 
1.70 & 
1.45 & 
1.06 & 
0.59 &
0.26 \\ 
$0.3$ & 
0.68 & 
0.95 & 
1.21 & 
1.38 & 
1.44 & 
1.38 & 
1.21 & 
0.95 & 
0.68 &
0.55 \\ 
$0.1$ & 
0.89 & 
0.96 & 
1.05 & 
1.11 & 
1.14 & 
1.11 & 
1.05 & 
0.96 & 
0.89 &
0.86 \\ 
$-0.3$ & 
1.23 & 
1.12 & 
0.94 & 
0.74 & 
0.65 & 
0.74 & 
0.94 & 
1.12 & 
1.23 &
1.26 \\ 
$-10.0$ & 
0.71 & 
1.11 & 
1.10 & 
0.69 & 
0.12 & 
0.69 & 
1.10 & 
1.11 & 
0.71 &
0.28 \\ \hline \hline 
\end{tabular}
\end{table}

\end{document}